\documentclass[twocolumn]{article}
\usepackage{usenix,epsfig,url}
\begin{document}

\date{}

\title{\Large \bf A Digital Preservation Network Appliance Based on OpenBSD}

\author{
David S.\ H.\ Rosenthal \\
{\em Stanford University Libraries} \\
{\em Stanford, CA 94305}\\
{\normalsize \url{http://www.lockss.org}} }

\maketitle

\thispagestyle{empty}

\subsection*{Abstract}

The LOCKSS program has developed and deployed in a world-wide test
a system for preserving access to academic journals published on
the Web.
The fundamental problem for any digital preservation system is
that it must be affordable for the long term.
To reduce the cost of ownership,
the LOCKSS system uses generic PC hardware, open source software,
and peer-to-peer technology.
It is packaged as a ``network appliance'', a single-function box that
can be connected to the Internet,
configured and left alone to do its job with minimal monitoring or
administration.
The first version of this system was based on a Linux boot floppy.
After three years of testing it was replaced by a second version,
based on OpenBSD and booting from CD-ROM.

We focus in this paper on the design,
implementation and deployment of
a network appliance based on an open source operating system.
We provide an overview of the LOCKSS application and describe the
experience of deploying and supporting its first version.
We list the requirements we took from this to drive the
design of the second version,
describe how we satisfied them in the OpenBSD environment,
and report on the initial deployment of this second version of
the appliance.

\section{\label{sec:introduction}Introduction}

The LOCKSS\footnote{LOCKSS is a trademark of Stanford University.}
(Lots Of Copies Keep Stuff Safe) program has
developed and deployed test versions of
a system for preserving access to academic journals published on
the Web.
The fundamental problem for any digital preservation system is
that it must be affordable for the long term.
To reduce the cost of ownership,
the LOCKSS system uses generic PC hardware, open source software,
and peer-to-peer technology.
It is packaged as a ``network appliance'', a single-function box that
can be connected to the Internet,
configured and left alone to do its job with minimal monitoring or
administration.
The appliance has to operate,
exposed to the Internet,
in environments lacking skilled
system administrators,
without imposing large administrative costs to install,
maintain or upgrade it.

The first version was based on a boot-floppy distribution of Linux.
After three years of testing at over 50 libraries world-wide,
this appliance level of the system was replaced by a second version,
based on a modified version of the OpenBSD install CD-ROM.
It was deployed to the test systems around the world in January 2003.
The application levels of the system have also been redesigned and
reimplemented from scratch;
deployment of this new implementation started in May 2003.

We focus in this paper on the design,  implementation and deployment of
a network appliance based on an open source operating system.
The goals and overall architecture of the LOCKSS system~\cite{Rosenthal2000},
the redesign of the protocol by which the peers
communicate~\cite{Maniatis2003lockssSOSP} and its economic
underpinnings~\cite{Rosenthal2003} are covered elsewhere.

We provide an overview of the application the network appliance is designed to
support.
We describe the experience of deploying and supporting its
first version.
We list the requirements we took from this as a basis for the
design of the second version,
describe how we satisfied them in the OpenBSD environment,
and report on the initial deployment of this second version of
the appliance.

\section{\label{sec:project}The LOCKSS Program}

Scientific communication has transitioned to the Web.
In particular,
much peer-reviewed science now appears only in e-journal
form~\cite{Keller2003}.
Academic journals are funded by university and other librarians
paying institutional subscription rates.
These librarians consider it part of their job to preserve access to the
record of science for future generations.
The transition to the Web has meant a transition from a purchase
model,
in which librarians buy and own a copy of the journal,
to a rental model,
in which librarians rent access to the publisher's copy.
Year-by-year rental provides no guarantee of future access,
and librarians fear the worst.
Publishers are motivated to allay these fears
and persuade libraries to switch to electronic-only subscriptions,
which save the publishers money.

The LOCKSS program is implementing
the purchase model for the Web,
providing tools librarians can use to take custody of,
and preserve access to,
web-published journals.
The tools allow libraries to run persistent web caches that:

\begin{itemize}

\item \emph{collect} material by crawling the e-journal Web sites,

\item \emph{distribute} material by acting like a proxy cache to make it
seem to a library's readers that web pages are still
available at their original URL,
even if they are no longer available there from the original
publisher~\cite{Spinellis2003}.

\item \emph{preserve} material by cooperating with other library's caches
in a peer-to-peer network to detect and repair damage.

\end{itemize}

The LOCKSS cache is implemented as a single daemon process written in Java.
It includes a specialized web crawler,
the Jetty~\cite{Jetty} web server to provide both a Web proxy
and an administration user interface,
and the peer communication protocol.

We believe that the major threats to digital preservation are economic
rather than technological; library budgets,
especially for preservation,
are never adequate~\cite{ARLstats}.
If LOCKSS is to succeed in preserving access for future generations,
libraries must be able to afford the system in the long term,
through the inevitable ups and downs of their budgets.
The goal of the LOCKSS program is to reduce the economic risks by
spreading the total cost of preservation across many independent
budgets,
minimizing the impact on each individual budget,
and lowering the economic barrier to entry.

Cost reduction has,
therefore,
been a major focus of the program from its inception.
Both the LOCKSS daemon and its operating system platform
are free and open source,
and use generic PC hardware to reduce hard costs as much as possible.

Soft costs,
especially support and system administration,
can easily dominate the hard costs.
Maintaining availability of Internet services over the long term is
very difficult.
Even at sites with expensive, skilled professional system adminstrators
between 20-50\% of outages are caused by
operator error~\cite{oppenheimer2003}.

Although it would be possible to include the LOCKSS daemon among a
variety of services on a large server,
the very long time horizons and slow operation of the system are
unlike the other services that it might be running.

Machine boundaries can provide effective fault isolation.
This encourages our belief that the overall reliability and cost of the
system can be improved by packaging the daemon,  its system infrastructure,
and the hardware it needs into a network appliance.
By reducing the economic barrier to entry as far as possible we hope to
broaden the base of libraries engaged in digital preservation activities.
On the other hand,
we expect many larger libraries will run the LOCKSS daemon without
its appliance platform.

Our network appliance design is intended to reduce both the cost and the risk
of system administration by identifying the expensive and risky operations
involved,
and either eliminating or automating them.
Our top candidates are installation,
upgrade,
and recovery from compromise.
We have succeeded in almost completely eliminating system administrator
involvement in all three.
We believe our ideas and experience in this area could
be helpful to others.

\section{\label{sec:lessons}Lessons}

The first version of the LOCKSS appliance~\cite{Rosenthal2000}
was based on the Linux Router Project (LRP) platform~\cite{LinuxRouter},
a boot floppy containing a minimal but functional Linux
system in a RAMdisk.
For our application,
it was capable of downloading and installing into a temporary
file system the LOCKSS daemon and the software on which it depended,
such as the Java Virtual Machine (JVM),
that would not fit on the floppy,

To begin running the appliance,
the host institution downloaded and ran a Windows program
that formatted,
wrote and checked a generic version of the floppy.
When a generic PC was booted from this floppy,
it asked the operator for the necessary configuration
information then personalized the floppy,
partitioned the disk and created the necessary file systems on it.

Booting the personalized floppy ran the normal LRP boot sequence then
downloaded the necessary LOCKSS software,
installed it in a temporary file system,
and finally invoked the JVM to run the daemon.

The system evolved over about 3 years of testing to run
at over 50 libraries worldwide and was generally sucessful
in requiring neither great skill nor much attention from
the host institution.
This taught us many valuable lessons.
The most important of these was that
running the system exclusively from write-locked media (and
from software verified against hashes on write-locked
media) greatly simplifies two critical tasks:
installing and configuring a new system,
and recovering from a compromise.

In unskilled hands the process of
installing and configuring a Unix system to be adequately secure
is a daunting and error-prone task.
Running an almost completely pre-configured system from write-locked
media obviates almost all the effort and risk.

Restoring a compromised system is a daunting and error-prone task
even in skilled hands.
A simple reboot is all that is needed to restore the LOCKSS appliance to
a known state.
If the compromise damaged or even completely destroyed the stored content,
the normal peer-to-peer communication will compare the damaged content with
the content at other peers and repair it from them or the publisher.

It may take many weeks to completely recover a destroyed peer.
However,
once the damage has been detected any potentially damaged and not yet
repaired content actually requested by readers
can be obtained from other peers.
Old academic papers are infrequently accessed~\cite{Anderson2001},
so the load of these requests is not significant.

Other important lessons we learned included:

\begin{itemize}

\item The floppy disks,
used only during boot,
were remarkably reliable.
We had very few media problems.

\item Operators did not always pay attention to instructions to
write-lock their floppy disks.

\item Squeezing the software we needed into the limited confines of even
a 1.68MB-formatted floppy,  and working with non-standard floppy formats,
was too time-consuming and error-prone for our small team.

\item In the interests of security,
MD5 hashes of the downloaded platform software were stored on the floppy.
It was thus necessary to persuade each test site to create a new
boot floppy before it could run a newly-released version
of the platform software,
for example to patch a vulnerability.
This had its bright side,
in that we were sure their configuration was consistent,
but it was an operations nightmare and very time-consuming.

\item Partly because of this inefficient upgrade path, the
small number of security vulnerabilities discovered
in our environment during the test consumed a totally
disproportionate amount of the team's efforts.
This was compounded by the way they tended to occur during
holidays.

\item A second reason for the difficulty in responding to
security vulnerabilities was the need to rebuild
the distribution after patching it.  LRP's build process
was never very robust and our modifications were so
extensive that building it became too fragile a process
to manage under the time pressure of a security incident.

\item LOCKSS caches need static,
globally routable IP addresses and communicate via UDP.
They can be run behind firewalls or network address translation (NAT)
boxes but doing so involves negotiation with the host institution's
network administrators.
If they are run outside the firewall,
a different negotiation with the network administrators may be required.
Some sites insist on security audits before installation.
The LOCKSS appliance needs a simple,
easy-to-explain,
and convincing security story.

\end{itemize}

\section{\label{sec:requirements}Requirements}

From this experience we developed requirements for the
second version of the system:

\begin{itemize}

\item Use only generic PC hardware.

\item Install the operating system,
application and all other software afresh in
a newly-created evanescent filesystem on every boot.

\item Even if the system is compromised,
there must be no place a Trojan horse could be hidden.
All file systems that persist across reboots must be mounted
with \emph{noexec}, \emph{nosuid} and \emph{nodev} options.

\item All software,
including the daemon,
must either run directly from read-only media
or from packages whose signatures are verified by
software running directly from read-only media before
being installed into evanescent file systems.

\item The public keys used to verify the signatures,
and thus permitted to sign software,
must be stored on write-locked media.

\item Each package should carry multiple signatures.
It must be impossible for revocation of a single key
to cause the system to fail.

\item The set of keys trusted to sign packages must be under the
control of the appliance operator.

\item Media containing system configuration data or keys
must be write-locked while the network is up.

\item Non-writability of media must be tested, not assumed.

\item No process sending or receiving network data may run as root.

\item The system must walk the user through the configuration
process and test that the supplied values work before
accepting them.

\item Upgrading the system should require only a reboot.

\item The response to a newly-discovered vulnerability should
be simple and rapid,  both to remove the vulnerability
and repair any damage to a compromised appliance's state.

\item The system must be easy for a small team to support.
Our criteria for this were:

\begin{itemize}

\item Use a major OS distribution with security as its primary focus;
don't use a minor distribution with a limited base of support such
as LRP.

\item Don't use a floppy-based distribution such as PicoBSD~\cite{PicoBSD};
squeezing GnuPG and other things we need onto a floppy is not feasible,
and the industry is phasing out floppy disks.

\item Don't change the OS build-from-source process.

\item Maintain a minimal footprint in the OS source.

\item Build the entire environment from scratch automatically
every night.

\end{itemize}

\end{itemize}

\section{\label{sec:trustModel}Trust Model}

We assume that libraries are capable of maintaining the physical
security of their LOCKSS appliances.
This leads us to trust to some extent the content of write-locked
media in the machine's drives;
an attacker capable of replacing the CD from which the machine
boots can obviously damage the machine's content.

We limit this trust to the boot image and a few other files
on the CD for which there wasn't space in the image
(Section~\ref{sec:background}).
All other software must carry a valid signature from at least one of
a set of keys which are trusted by the system's administrator and not
known to be revoked.  The trusted software consists of the kernel and
the utilities needed to perform the key revocation check,
signature validation and software installation.

This technique has
similarities to the Trusted Computing Platform Architecture
(TCPA)~\cite{TCPA}.
Our goals,
however are quite different.
TCPA is intended to enable programs which do not trust the administrator
of the system on which they are running to verify that
the system's integrity is
attested to by keys that the program does trust.
LOCKSS peers do not trust each other and do not run third-party software;
the integrity of the software on a LOCKSS cache is of purely local interest.
Thus our,
much weaker,
goal is to assure the administrator of the system that,
at least immediately after a reboot,
it is running only software whose integrity is attested to by
keys that the adminstrator trusts.

Lacking the hardware and BIOS support of TCPA,
we cannot fully achieve this goal.
A remote root compromise could in some circumstances
allow the attacker to modify the
system's BIOS and thereby disable
the signature verification process.
This in turn might allow spurious software to persist across reboots.
To minimize this and other risks of a remote root compromise,
we run with OpenBSD's \emph{kern.securelevel} variable set to 2,
the most restrictive possible,
and we remove the debugger from our kernel.

\section{\label{sec:background}Background}

These requirements led us to a design based on adapting the OpenBSD
install CD.
The design of this bootable CD is a series of layers,
which we describe from the outside in:

\begin{itemize}

\item Booting a PC from a CD requires that the image of a
2.88MB ``boot floppy'' be present on the CD.

\item In the case of OpenBSD,
this ``boot floppy'' contains an FFS file system and the
\emph{biosboot} program, which the PC's BIOS locates and starts.

\item The file system contains a compressed kernel image,
which \emph{biosboot} loads into memory and executes.

\item Part of the kernel's data space is a RAMdisk image containing an FFS file
system 1.7MB big,
which is mounted as the root file system.

\item This file system contains skeletal system directories such as \emph{/etc}
and \emph{/dev}.
It also contains a single \emph{crunched} binary which implements
all the commands needed for the installation.

\end{itemize}

The kernel starts up in the normal way,
but it runs a special \emph{/etc/rc} script that walks the user through the
installation process.
The shell and all the other commands needed to run this script are in
the crunched binary.

The crunched binary is constructed from the regular source tree
by a pair of tools called \emph{crunchgen} and \emph{crunchide}.
Crunchgen works 
from a list of commands by
locating the Makefile for each command in the source tree,
and writing a new Makefile,
a set of stub programs and a top-level \emph{main()} that calls the
stub chosen by \emph{argv[0]}.
The new Makefile builds the appropriate set of object files
from the command sources,
then links each command and its stub into an intermediate file that
is processed by crunchide to hide all its global symbols
except those of the stub.
Finally,
the intermediate files are linked with the generated \emph{main()}
into the crunched binary.

\section{\label{sec:design}Design}

Our approach was to replace the \emph{/etc/rc} script run by the minimal
``boot floppy'' environment on the install CD with a slightly
augmented regular system's \emph{/etc/rc} which:

\begin{itemize}

\item Establishes swap space.
\item Confirms that the configuration floppy is not writable.
\item Creates an evanescent file system in the swap space;
redirecting the system directories into it via symlinks.
\item Validates the signatures on the
necessary software packages.
\item Installs the packages into the evanescent file system
(via the symlinks).
\item Installs the network configuration information from the floppy.
\item Runs only the essential services:
\begin{itemize}
\item The SSH daemon~\cite{Ylonen1996} for remote administration,
with privilege separation~\cite{Provos2002}
\item The LOCKSS daemon inside the JVM run as an unprivileged user.
\item The appliance does not run sendmail.  Outbound mail is sent by ssmtp,
a minimal SMTP client.  Inbound mail is not accepted.
\end{itemize}
\item Installs a \emph{crontab} entry that implements the automatic
package update mechanism.

\end{itemize}

The package update mechanism at intervals chooses at random
from a list of download servers.
As an unprivileged user it checks that server for new packages and
signatures and downloads any found.
It caches them on the hard disk.
They will be used at the next reboot if their signatures are valid.
Although a compromise could allow these cached packages to be modified,
doing so would invalidate their signatures.
An invalid signature causes the boot process to ignore that version of
the package and revert to an earlier version,
probably the one on the CD.

\section{\label{sec:build}Build Process}

The implementation
resides entirely in directory hierarchies below
\emph{/usr/src/distrib/i386/lockss/}.
except for:

\begin{itemize}
\item The LOCKSS kernel config file.
\item Three directories in \emph{/usr/src/distrib/special/}:
\begin{itemize}
\item A skeleton implementation of \emph{host}, to avoid the full glory of the
BIND implementation.
\item A skeleton implementation of \emph{sudo}, capable only of being used by
root to give up privileges.
This allows us to avoid root processes accessing the network during the boot
sequence.
\item An altered Makefile for \emph{init},  needed to get it to go multi-user
by default.
\end{itemize}
\item An additional entry in \emph{/usr/src/distrib/i386/Makefile} that
builds the LOCKSS CD.
\end{itemize}

Our goal of maintaining a small footprint in the
distribution source has been met.

Our build script starts by checking the patch branch out from
OpenBSD's AnonCVS~\cite{Cranor1999} service into a temporary
build tree,
updating our copy of the ports tree,
then checking the implementation of the LOCKSS appliance platform out
from our CVS server and copying it into place in the build tree.

The build process then patches the
kernel source to install one additional driver,
for \emph{swdt},
a kernel-based software watchdog that reboots the system if a
user-level daemon fails to reset the watchdog at least every 30 seconds.
It can sometimes recover a system that has locked up.

The build process is essentially the OpenBSD install CD build process.
It results in a bootable CD very similar to OpenBSD's,
but containing in addition to the basic OpenBSD packages a set of packages
from the ports tree,
including the JVM,
the Red Hat emulator needed to run it,
ssmtp and some libraries they depend on.
It also contains a set of LOCKSS packages containing the daemon and its
environment.
Installed on the CD itself is GnuPG~\cite{GnuPG},
and the shared libraries it needs to run.

Distributing new CD images,
having administrators burn them and use them to reboot their systems
is too time-consuming and expensive to fix security vulnerabilities or
for routine upgrades to the application.
The \emph{crontab} entry described above can download packages
and cache them on the disk for use at the next reboot.
To avoid mistakes,
there is no separate build process for packages distributed in this way.
An upgrade package is built by building an entire CD image,
extracting the package from it and signing it.
The package and its signature are placed on the download
servers,
and mail is sent to the sites asking them to:

\begin{itemize}

\item log in as root

\item execute a script that checks for new packages

\item reboot

\end{itemize}

\section{\label{sec:implementation}Implementation}

The implementation consists almost entirely of three shell scripts,
residing in the RAMdisk image in the CD's "boot floppy",
which are executed during the system boot process.
The work they do means that our appliance takes a rather long time to
boot,
but in the field of digital preservation speed is not a requirement
(see \ref{sec:experiencePerformance}).
The scripts are:
\begin{itemize}

\item \emph{/etc/rc.0.lockss}, which is executed very early in \emph{/etc/rc},
before swap is enabled;

\item \emph{/etc/rc.1.lockss}, which is executed later in \emph{/etc/rc},
once the file systems are mounted;

\item and \emph{/etc/lockss.start}, which is executed at the end of
\emph{/etc/rc.local}.

\end{itemize}

The implementation also uses GnuPG
which (together with the libraries it needs) is accessed
directly from a directory on the CD,
to ensure that its testimony as to the trustworthiness of
the packages to be installed can be trusted.
The keys used in this process are on the write-locked
floppy and are thus under the control
of the appliance operator,
who can add or delete keys,
and also add signatures to the floppy.
This is an important point;
forcing our users to trust our keys would allow us to shut the
system down if we chose.

The remaining part of the implementation consists of some changes to
the list of files to be "crunched"
together to form the command binary in the minimal system in the CD's
"boot floppy".

\subsection{\label{sec:implementation0}/etc/rc.0.lockss}

This script is run just before swap is turned on to ensure that there
is swap space available.  It examines the available hard disks and,
if they have not yet been partitioned appropriately,  asks permission
then does so.

The first hard disk is partitioned with about a gigabyte of swap space;
the remainder and all other disks are used as file systems to contain
preserved content.
It creates an appropriate \emph{/etc/fstab} describing them and the
MFS~\cite{Snyder1990} file system that will be created later in the swap space.

\subsection{\label{sec:implementation1}/etc/rc.1.lockss}

\begin{figure}
\centerline{\includegraphics{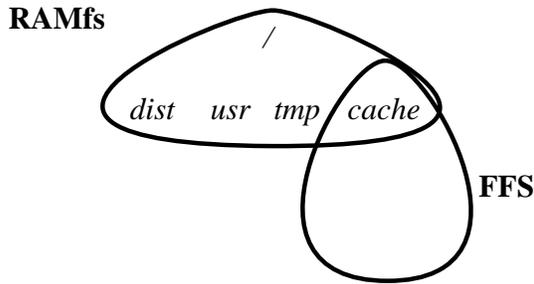}}
\caption{The file name space as \emph{/etc/rc.1.lockss} starts.}
\label{fig:Phase1}
\end{figure}

This script implements the bulk of the LOCKSS "platform".
We describe it in narrative form.
The file name space as the script starts is defined by the \emph{/etc/fstab}
written by \emph{/etc/rc.0.lockss}.
It is shown in Figure~\ref{fig:Phase1}.

It attempts to mount the CD the system was booted from and,  if it
cannot,  calls for help.  It is not unknown to find that the
kernel autoconfiguration process has failed to recognize the CD
from which it was booted.

It then mounts the MFS file system on \emph{/dist}.
At this point the \emph{/tmp} directory is still where it was on boot,
in the RAMdisk,
and it is about to run out of space.
The script relocates \emph{/tmp} into the newly-created MFS by
copying its content to \emph{/dist/tmp} and replacing it with a symlink
there.

It checks to see if there is a floppy in the drive.  If not,  it runs
the configuration process described below.  If there is one but it
is write-enabled,  the script refuses to proceed until it is write-locked.
If there is a write-locked floppy,  any configuration information it
contains supercedes any configuration information on the CD.
The GnuPG public keyring is initialized from the floppy.

It then checks all the packages it can find on the package path and
writes a script that,
when later executed,
will for each necessary package install
the latest available version which has at least one valid signature.
The package path consists of the floppy,
specific directories on the CD,
and the cache of downloaded packages on the hard disk.
Note that during this process the entire running
environment consists of the RAMdisk image from the CD,
GnuPG (being run directly from the CD),
and configuration data from a floppy that is known to be write-locked.
The checking process is as follows:

\begin{itemize}

\item The network is bought up using the configuration information but
without any daemons running.

\item GnuPG is invoked as an unprivileged user
to do a key revocation check on its keyring.

\item The network is shut down.


\item The valid-MD5 list is initialized to be empty.

\item All detached signatures for any MD5 files found in the package path
are checked.
If at least one valid signature by a key that is not known to be revoked is 
found,  that file's list of MD5s is added to the valid-MD5 list.

\item The install-package script is initialized to be null.

\item The package path is searched for versions of the necessary packages.
The MD5 of each version found is computed
and compared against the valid-MD5 list.  If a match is found, and
a command to install a lower version number of the package is already in the
install-package script,  it is deleted.  Then a command to install
this version is added to the script.

\item If no unrevoked signature validates the MD5 of a package that must be
installed,
the boot process is aborted and the system
enters ``hunker-down'' mode.
It,
and its preserved content,
are inaccessible from the network and thus should be safe for some
time.
The administrator will need to add new signatures to the floppy
before resuming normal operation.

\end{itemize}

The script next prepares for package installation by creating copies
of the system directories under \emph{/dist}
and replacing the originals with symlinks pointing to the copies.

The system then executes the install-package script it wrote.
This first installs the OpenBSD base "packages" into directories in 
the MFS under \emph{/dist},
then replaces the ``boot-floppy'' system directories with symlinks
to the installed versions in \emph{/dist}.
Now the file name space approximates a properly installed OpenBSD system,
and the install script continues to install a chosen set of real
packages from the ports tree,
and a small number of LOCKSS packages.
Although any of these packages may come from any of the directories
in the package path,  each is known to have had its MD5 signed by at
least one unrevoked key trusted by the system's administrator.

\begin{figure}
\centerline{\includegraphics{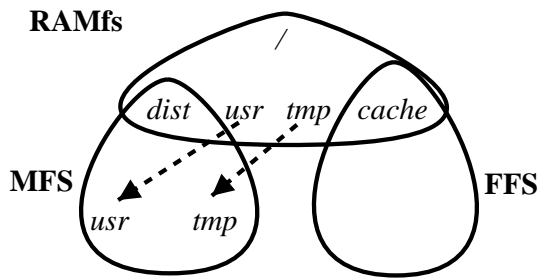}}
\caption{The file name space as \emph{/etc/rc.1.lockss} ends.}
\label{fig:Phase2}
\end{figure}

Finally, the script unmounts everything it mounted and returns to the
\emph{/etc/rc} script.
The file name space at this point is shown in Figure~\ref{fig:Phase2}.

\subsection{\label{sec:implementationStart}/etc/lockss.start}

This script is run by \emph{/etc/rc.local} just before the end of the system
boot process,  after the few system daemons we have not disabled
have been started.
It enables the watchdog,
checks for the required LOCKSS daemon configuration files and,
if it finds them,
invokes the JVM to run the LOCKSS daemon.

\subsection{\label{sec:implementationConfiguration}Configuration}

LOCKSS peers use their IP address as their long-term identity and thus,
at present,
do not support DHCP.
If no configuration information is available,
early in \emph{/etc/rc.1.lockss} a text dialog is used to acquire
an IP address,
netmask,
gateway,
DNS servers,
root password etc.

The supplied values are validated by using them to bring up the
network interface temporarily and,
as an unprivileged user,
perform some simple tests such as a DNS lookup.
If the tests work,
the network interface is shut down.
The user is asked to write-enable the floppy,
and the configuration is written to the floppy as a text file.
New SSH host keys are generated and stored on the floppy,
together with the initial GnuPG public keyring.
The user is then asked to write-lock the floppy.
The system verifies that it is write-locked before proceeding.

\section{\label{sec:experiencePerformance}Performance}

One performance issue that might make these techniques unsuitable for
other applications is that it takes the system some time to boot.
On a 400MHz PC,
somewhat slower than those our test sites typically use,
with a 40X IDE CD drive,
the boot sequence takes about 320s from reset to login prompt.
This includes about 30s checking signatures and hashes,
about 25s installing the OpenBSD base packages,
and about 180s installing about 96MB of other packages.

If our caches reboot once every 60 days this would
represent about 0.006\% downtime,
a reasonable price to pay for low administrative costs.
Although we don't have accurate data,
we believe our test caches reboot significantly less often than this.
Obviously,
reducing the size of the packages to be installed would reduce the
impact of rebooting on availability,
and there is much scope for doing so.

Once the system is up,
its performance is indistinguishable from a similar system
running a more conventional configuration of OpenBSD.

A more critical performance measure is the time taken to
respond to a security vulnerability.
We simulated this by involving our test sites in a ``firedrill'',
in which we generated and distributed a ``patch'' that simply sent us
an e-mail confirming that the system had downloaded and
was using it.
The firedrill showed that building and distributing a patch is now far more
efficient than it was for the Linux floppy version.
With little effort on our or their part we were able to get 96\%
of the deployed systems upgraded within 48 hours of the start of the
firedrill.

\section{\label{sec:experienceDeployment}Deployment}

In general,
the design,
implementation and rollout of the new version to over 50 test
sites went smoothly.
Almost all our test sites have been trouble-free since
their upgrade in January 2003 from the floppy version.
This is a great tribute to the quality of OpenBSD.

The LOCKSS appliance has survived security audits by some fairly
demanding sites (e.g. CERN and LANL).

\section{\label{sec:experienceProblems}Problems}

We did encounter a number of problems.

Our initial design involved union mounting the MFS file system
over the root directory.
This didn't work;
it seemed to cause deadlock in the kernel.
We have not verified that the problem is still present;
working around it seemed preferable to diagnosing problems
that early in the boot sequence.

OpenBSD appears to provide no way to check whether a floppy is
write-locked or write-enabled without mounting it,
and seeing whether an attempt to write to it fails.
If it does fail,  the kernel generates alarming error messages
which the operator has to be told to ignore.
OpenBSD currently lacks FreeBSD's mechanism for selectively disabling
these messages.

We orignally wanted to use the system itself to burn the CDs.
Again, the operator has to ignore some alarming error messages,
so we put the idea aside.

The kernel autoconfiguration sometimes fails to recognize the CD
from which it was booted,
causing the attempt to mount the CD to fail,
and preventing the system installing software from it.
Power-cycling the machine seems to be the only cure.

The native 1.3 JVM for OpenBSD wasn't available,
so we run the Linux JVM under the RedHat emulator.
This works well but installing the emulator and the RPM package
slows the boot considerably and makes us nervous.

The RPM-based package install for the JVM insists on executing from a path
in \emph{/usr};
we have to specifically turn off \emph{noexec} during this
install and re-enable it afterwards.

We ask our test sites to download an image of each new version of
the CD as a \emph{.iso} file,
write a CD-R of it using some other machine,
and boot their cache machine from it.
The first time they tried this,
about 1 in 20 sites had problems of one kind or another that required
support from the team.
Examples are difficulty with Windows CD-writing software,
difficulty with boot device settings in the BIOS,
and bad CD-R media.

Despite our suggestion that they use second-hand machines,
our enthusiastic test sites often buy brand-new hardware.
This frequently has motherboard Ethernet chips which are too new for
the OpenBSD drivers to use.
We have accumulated a reserve of really old Ethernet cards to mail
out in such cases.

The security vulnerability firedrill demonstrated a need to
improve our process for making a patch available.
In particular,
we need to focus on making our test sites aware that when we say
``you need to reboot your system right now'' they need to pay
attention.
We hope that exercising this process regularly will make it
go more smoothly when it is needed ``for real'';
the 96\% success in 48 hours of the first drill is not good enough.

\section{\label{sec:relatedWork}Related Work}

Our work has similarities with KNOPPIX\cite{Knopper2000} and other
bootable Linux CDs used as demo and rescue systems.  They typically
run directly from the CD rather than,  as we do, installing system
packages into an evanescent file system.  We take a lot longer to
boot while we do the installation,  but we can validate the
signatures on the packages,  and thus use downloaded signed packages
to upgrade the system on the CD.

Sun's Cobalt product line~\cite{Cobalt} is
an excellent example of packaging Linux into an network appliance that
works well with limited administration.
Our automatic software upgrade mechanism was inspired by Cobalt's,
which has automatic notification of availability but manual installation.

\section{\label{sec:futureWork}Future Work}

Work on the platform is on hold for a while;
while we focus on using it to deploy a completely re-written version of the
daemon.
When we get back to it,
our to-do list includes in descending order of priority:

\begin{itemize}

\item supporting DHCP, NAT, and machines behind firewalls,
\item using native Java,
\item supporting USB storage devices for configuration,
\item building a KNOPPIX-like LOCKSS demo mode,
\item burning CDs from the system itself to allow packages and configuration
data to be stored on the CD,
\item revisiting the idea of union mounting MFS over the root.

\end{itemize}

There is also the possibility of extending this work to interesting and useful
areas that aren't directly related to the mission of LOCKSS.
Extensions to the stackable Open Source BIOS~\cite{agnew2003},
which is based on LinuxBIOS~\cite{minnich2000},
could provide a trusted boot sequence on which we could base our verification
of the higher levels of the system.
Combining this with a physically write-protected USB ``dongle'' containing
the keys could provide almost all the capabilities of TCPA for Open Source
systems except tamper-resistance and secret key protection.

\section{\label{sec:conclusion}Conclusion}

In many ways,
OpenBSD has proven an excellent basis for our network appliance.
It was easy to adapt the install CD to our purposes.
The carefully created default configuration is an excellent starting point
for further restrictions.
The use of privilege separation in the SSH daemon reduces the risk of
allowing adminstrative access via the network.
OpenBSD 3.3's addition of Stack Smashing Protection~\cite{ProPolice}
adds further protection.
AnonCVS access to the patch branch and the ports tree has allowed us
to build an up-to-date system from scratch every night,
materially reducing the load on a small support team,
especially when responding to new vulnerabilities.
The OpenBSD build process is well-structured and allowed us to add
a new CD image with little effort,
both initially and as the base distribution evolved from 3.1 to 3.2 then 3.3.

We have added to OpenBSD a set of capabilities that allow it to serve
quite satisfactorily as a network appliance with unskilled administrators,
if only for an application that can tolerate extended reboot times.
These include:

\begin{itemize}

\item Running the system entirely from evanescent file systems
re-created from write-locked media at boot time,
with no ability to execute code from a persistent file system.

\item Verifying the signatures on all software during the boot
process.

\item Implementing a semi-automatic patch distribution mechanism
for packages and their signatures.

\end{itemize}

There are,
however some deficiencies in OpenBSD that still cause significant problems
as we deploy it as an appliance.
Those causing the most support load are:

\begin{itemize}

\item The kernel produces many scary-looking error messages in
non-error situations.
\item The kernel does not reliably recognize low-cost IDE CD drives.
\item The NIC drivers are sometimes unable to recognize or use
leading-edge hardware.

\end{itemize}

There is never a perfect choice of platform for an application;
all choices have advantages and disadvantages.
OpenBSD has served us as well as we expected,
and we hope that others implementing network appliances will
find our experiences useful,
whether they agree with our choice or not.

\section{\label{sec:acknowledgements}Acknowledgements}

This material is based upon work supported
by the National Science Foundation under Grant No. 9907296, however any
opinions, findings, and conclusions or recommendations expressed in this
material are those of the author and do not necessarily reflect the
views of the National Science Foundation. 

The LOCKSS program is grateful for support from the National Science
Foundation,  the Andrew W. Mellon Foundation, Sun Microsystems
Laboratories, and Stanford Libraries.

Special thanks are due to Mark Seiden, who made major contributions to
both versions of the platform and especially to the signature verification
process.
Thanks are due also to our long-suffering beta sites,
and to the LOCKSS engineering team of
Tom Robertson,
Tom Lipkis,
Claire Griffin,
and Emil Aalto.
Our web crawler is adapted from an original by James Gosling.

Vicky Reich has made the LOCKSS program possible.

The anonymous BSDcon reviewers and Todd Miller,
the paper shepherd,
provided many pertinent and helpful comments.

Finally,
the thanks of the entire LOCKSS team go to everyone who has contributed to
OpenBSD,
GnuPG,
the JVMs,
and the Jetty web server.

\section{\label{sec:availability}Availability}

The source for the entire LOCKSS system,  including the appliance platform
described above, carries BSD-style Open Source
licenses and is available from the LOCKSS project at SourceForge.

\bibliographystyle{plain}
\bibliography{../common/bibliography}

\end{document}